\documentclass[%
 aip,
 amsmath,amssymb,
 reprint,%
]{revtex4-1}

\usepackage{graphicx}
\usepackage{dcolumn}
\usepackage{bm}
\usepackage[normalem]{ulem}

\usepackage[utf8]{inputenc}
\usepackage[T1]{fontenc}
\usepackage{mathptmx}
\usepackage{etoolbox}
\usepackage{xcolor}
\usepackage[colorlinks=true, linkcolor=blue, citecolor=blue, urlcolor=blue]{hyperref}

\makeatletter
\def\@email#1#2{%
  \patchcmd{\titleblock@produce}
    {\frontmatter@RRAPformat}
    {\frontmatter@RRAPformat{\produce@RRAP{*#1: \href{mailto:#2}{#2}}}\frontmatter@RRAPformat}
    {}{}
}%
\makeatother

\begin{document}

\title{Atomistic Simulations of Oxide-Water Interfaces using Machine Learning Potentials}

\author{Jan Elsner*}
\affiliation{Lehrstuhl f\"ur Theoretische Chemie II, Ruhr-Universit\"at Bochum, 44780 Bochum, Germany}
\affiliation{Research Center Chemical Sciences and Sustainability, Research Alliance Ruhr, 44780 Bochum, Germany}
\author{K. Nikolas Lausch}
\affiliation{Lehrstuhl f\"ur Theoretische Chemie II, Ruhr-Universit\"at Bochum, 44780 Bochum, Germany}
\affiliation{Research Center Chemical Sciences and Sustainability, Research Alliance Ruhr, 44780 Bochum, Germany}
\author{J\"{o}rg Behler}
\affiliation{Lehrstuhl f\"ur Theoretische Chemie II, Ruhr-Universit\"at Bochum, 44780 Bochum, Germany}
\affiliation{Research Center Chemical Sciences and Sustainability, Research Alliance Ruhr, 44780 Bochum, Germany}

\makeatletter
\@email{Corresponding author}{Jan.Elsner@ruhr-uni-bochum.de}
\makeatother

\date{\today}

\begin{abstract}
Oxide-water interfaces govern a wide range of physical and chemical processes fundamental to many fields like catalysis, geochemistry, corrosion, electrochemistry, and sensor technology. Near solid oxide surfaces, water behaves differently than in the bulk, exhibiting pronounced structuring and increased reactivity, typically requiring \textit{ab initio}-level accuracy for reliable modeling.
However, explicit \textit{ab initio} calculations are often computationally prohibitive, especially if large system sizes and long simulation time scales are required. 
By ``learning'' the potential energy surface (PES) from data obtained from electronic structure calculations, machine learning potentials (MLPs) have emerged as transformative tools, enabling simulations with \textit{ab initio} accuracy at dramatically reduced computational expense. 
Here, we provide an overview of recent progress in the application of MLPs to atomistic simulations of oxide-water interfaces. Specifically, we review insights that have been gained into the reactivity of interfacial systems involving the dissociation and recombination of water molecules, proton transfer processes between the solvent and the surface and the dynamic nature of aqueous oxide surfaces. Moreover, we discuss open challenges and future possible research directions in this rapidly evolving but challenging field.
\end{abstract}

\maketitle 


\section{Introduction}\label{sec:Introduction}

The chemistry and physics of oxide-water interfaces play a critical role in many environmental and technological processes. 
In nature, reactions at these interfaces drive weathering, 
shaping landscapes and contributing to soil formation\cite{al2003oxide}. 
Interfaces also serve as critical sites for microbial activity, where microorganisms mediate metal oxide dissolution\cite{brown1999metal}.
In technology, oxide-water interfaces are central to
catalysis and photocatalysis\cite{vedrine2019metal}, 
facilitate environmental remediation by aiding in the removal of contaminants from water\cite{banuelos2023oxide}, and play important roles in semiconductor manufacturing\cite{reinhardt2011handbook}, 
drug delivery systems\cite{vangijzegem2019magnetic} and sensing technology\cite{nunes2019metal}. Moreover, corrosion of metals and alloys, i.e., oxide formation, has a significant economic impact with an estimated global cost of about 2.5 trillion US\$ per year \cite{P7019}.

Compared to bulk solids and liquids, interfacial systems have proven far more challenging to characterize, both theoretically and experimentally.
At surfaces, the disruption of bulk structural and chemical continuity leads to changes in the electronic structure compared to the bulk, often resulting in complex surface reorganizations\cite{polo2019surface}. 
At oxide-water interfaces, 
additional complexity arises from the dynamic interplay of two distinct phases in contact.
Surface electrostatics, hydrogen bonding, dissociation and recombination of aqueous species and solvation effects shape interfacial dynamics and chemical interactions in ways not predictable from bulk properties. 
If solutes are present, as in heterogeneous catalysis, their behavior and reactivity are strongly influenced by interactions with both the solvent and the surface, which in turn are shaped by the restructuring of the water network at the interface\cite{bin2024chemistry}.
Moreover, the behaviour at these interfaces is highly sensitive to factors such as the nature of the specific oxide facet, surface defects, and thermodynamic conditions like temperature and pressure, as well as pH and external fields.

A variety of experimental techniques have been employed to study oxide-water interfaces, including scanning probe microscopies (e.g., Scanning Tunneling Microscopy (STM)\cite{wang2007electrochemical,feng2021application} and Atomic Force Microscopy (AFM)\cite{santos2016imaging, chen2022principles}), electron microscopies (e.g., environmental transmission electron microscopy (ETEM)\cite{raabe2012situ, jooss2016etem}, X-ray-based methods (e.g., X-ray reflectivity\cite{fenter2004mineral, park2005probing}, Crystal Truncation Rod (CTR)\cite{robinson1986crystal, trainor2002crystal} and X-ray absorption\cite{arai2001x, regan2001chemical}), and vibrational spectroscopies (e.g., Infrared (IR)\cite{al2003ft, andanson2010exploring}, Raman\cite{zhang2017water, li2017core}, Sum-Frequency Generation (SFG)\cite{shen1989surface, du1994vibrational, du1994surface, shen2006sum, tuladhar2017insights}).  
Although these approaches provide a wealth of information, their interpretation often relies on theoretical support.
For instance, identifying the atomic-scale processes responsible for specific features in experimental spectra can be challenging and benefits greatly from accompanying theoretical studies to provide a comprehensive picture\cite{gaigeot2012oxide, sulpizi2012silica,sulpizi2013vibrational}. Moreover, computer simulations enable predictions about systems under thermodynamic conditions that are difficult to achieve experimentally, or about systems not yet realized in the laboratory.

Theoretical insight into the atomistic structure and dynamics of interfacial systems can be achieved through molecular dynamics (MD) simulations, which evolve the positions of atoms over time by numerically integrating Newton's equations of motion\cite{frenkel2023understanding}, thus naturally including thermal fluctuations, which are particularly important in the liquid phase. 
Such simulations depend crucially on the accuracy of the high-dimensional potential energy surface (PES) governing the atomic interactions. 
To accurately account for the complex interfacial environment, including reactive events such as water dissociation and proton transfer (PT)~\cite{hass1998chemistry, hu2010proton, sato2015proton,hussain2017structure}, an \textit{ab initio} description incorporating the subtleties of the system's electronic structure is generally required. 
For this reason, density functional theory (DFT)-based \textit{ab initio} molecular dynamics (AIMD)\cite{marx2009ab} has been an indispensable theoretical tool for investigating oxide-water interfaces over the last two
decades\cite{bjorneholm2016water, wang2021investigations}.    
However, DFT-based simulations are computationally expensive and many processes relevant to oxide-water interfaces span time or length scales beyond the reach of AIMD, which is typically restricted to a few hundreds of atoms interacting for tens of~ps. 
Computational models of interfacial systems where water behaves like the bulk liquid far from the solid surface require many hundreds of water molecules\cite{natarajan2016neural, quaranta2017proton}, making even single-point DFT calculations prohibitively expensive for more advanced exchange-correlation functionals. Furthermore, the short simulation times accessible in AIMD simulations make it challenging to obtain statistically converged properties such as water density profiles,  PT reactions, diffusion coefficients and vibrational spectra.

These challenges are increasingly addressed by machine learning potentials (MLPs), which are trained to accurately reproduce the ab inito PES and can be evaluated at a greatly reduced computational cost. 
MLPs have revolutionized computational chemistry and materials science by enabling \textit{ab initio} accuracy in simulations of systems and phenomena spanning diverse length and time scales\cite{behler2016perspective, deringer2019machine, behler2021machine, P5793,P6102,P6112,P6131,P6158,P6631}, including aqueous systems\cite{omranpour2024perspective}, interfaces\cite{artrith2019machine, zhou2023machine, houimproving} and heterogeneous catalysis\cite{tang2024machine, omranpour2024machine, P6896}.
Beyond traditional MLP models limited to predicting energies and forces from structure, 
machine learning 
can also be used to predict
electronic structural information, for instance dipole moment and polarizability surfaces,
which are required for simulations of, e.g., vibrational spectra.  
Moreover, MLPs have been developed that go beyond the traditionally assumed locality of interatomic interactions\cite{ghasemi2015interatomic, xie2020incorporating, ko2021fourth, unke2021spookynet, shaidu2024incorporating}, enabling the study of systems where nonlocal effects become significant -- a consideration that may turn out to be important for oxide-water systems with doping or defects. 

In this mini-review, we provide an overview of the progress made over the last decade using MLPs to investigate oxide-water interfaces. We begin with an overview of commonly employed MLPs for this purpose.
Next, we review applications of MLPs to oxide-water interfaces. 
Unless otherwise noted, the MLPs discussed here were trained on custom datasets for a specific oxide-water system. A discussion of the recent development of pre-trained models is provided in Section~\ref{sec:Conclusions and Outlook}.
We do not attempt to detail specific training procedures, active learning schemes, or model validations, which may be highly specific to individual systems and methods. Rather, we aim to provide an overview of the broad range of applications enabled by MLPs in studies of oxide-water interfaces, covering insights into water dissociation and PT reactions, as well as surface processes, including the influence of defects and dynamic reconstructions. Subsequently, we highlight individual studies that, on top of using MLPs to drive MD simulations, use additional machine learning models to recover electronic structural information. This is an important step toward machine learning-based simulations capable of capturing all properties of interest, beyond just energies and forces.
Finally, we provide a brief discussion and outlook on the challenges and opportunities that lie ahead.

\section{Machine learning potentials}\label{sec:Machine learning potentials}

\label{sec:Classification of machine learning potentials}

\begin{figure*}[!ht]
\centering
\includegraphics[width=1.0\textwidth]{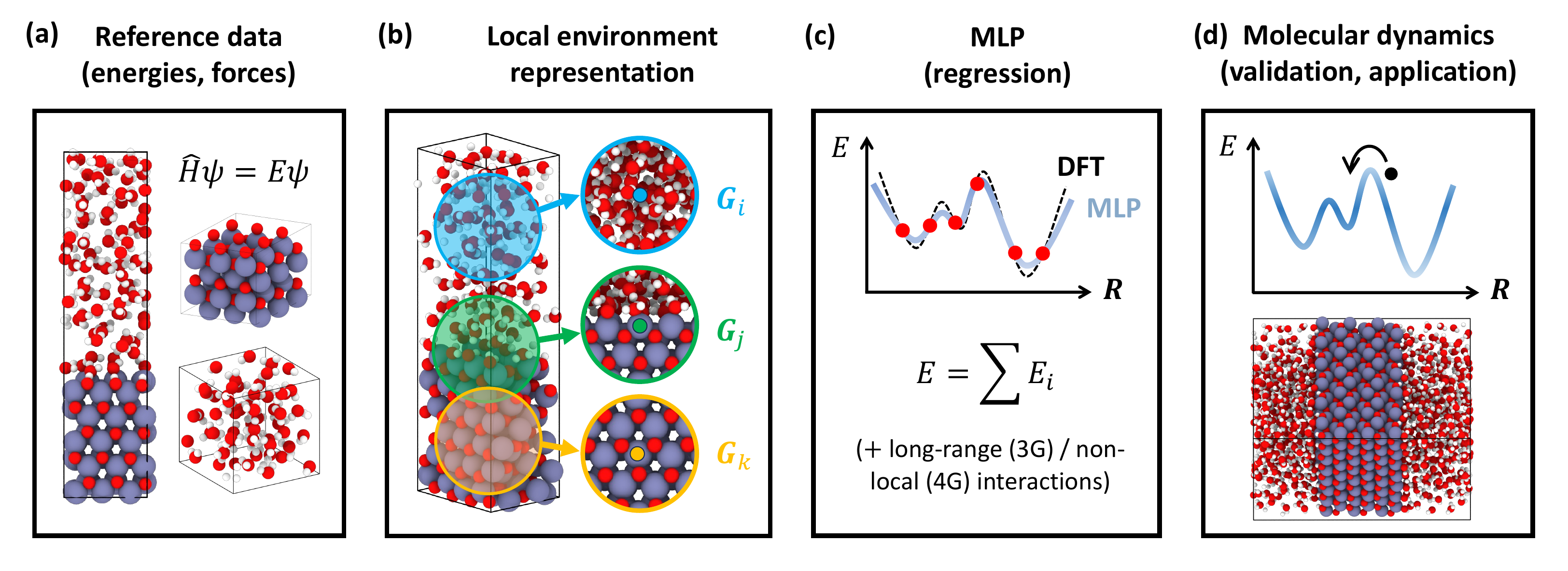}
  \caption{Schematic illustration of the key principles underlying the construction and application of MLPs for water-oxide interfaces. 
  (a) Quantum mechanical calculations (usually DFT) provide reference energies and forces for selected atomic configurations, which serve as training and testing data. 
  (b) The local environment (i.e., atoms within a cutoff radius) of each atom $i$ is mapped to a descriptor $\mathbf{G}_i$ that encodes the required invariances of the potential energy, $E$, to translation, rotation, and permutation of chemically equivalent atoms. Such descriptors may be predefined (e.g., HDNNPs and GAP), or learned during training (e.g., DeePMD and MPNNs). In MPNNs, iterative message passing operations allow $\mathbf{G}_i$ to be influenced by atoms outside the cutoff radius.
  (c) The MLP is trained to reproduce the reference energies and forces using supervised learning. In most architectures, the potential energy is expressed as a sum over atomic contributions, Eq.~\ref{eq:e_decom} (2G MLPs), though models may incorporate additional terms to capture long-range (3G MLPs) or nonlocal interactions (4G MLPs).
  (d) Once trained, MLPs enable MD simulations of systems much larger than those used for training, while retaining near-DFT accuracy.}
  \label{fig:MLP_fig_CPR2025}
\end{figure*}

MLPs seek to provide the accuracy of \textit{ab initio} methods at much reduced computational cost by approximating the PES -- the functional relation between the geometry and the potential energy of a system under the Born-Oppenheimer approximation -- using machine learning (ML) algorithms trained to accurate reference data obtained from quantum mechanical electronic structure calculations. By avoiding explicit treatment of electronic degrees of freedom and making use of efficient machine learning architectures, 
MLPs can be evaluated several orders of magnitude faster than the underlying reference method, typically introducing only very small energy errors of about 1\,meV\,/\,atom and force errors of about 100\,meV\,/\,\AA. An overview of the main principles of MLPs is shown in Figure~\ref{fig:MLP_fig_CPR2025}.

Nowadays, the development of MLPs has become a huge field of research, and in this section we provide only a brief overview of the available methods. Readers interested in further details on the wide range of methods, their properties, training and validation are referred to a large number of review articles dedicated to these topics \cite{behler2016perspective, deringer2019machine, behler2021machine, P5793,P6102,P6112,P6131,P6158,P6631,P6346,P6548,jacobs_practical_2025}. Here, we will follow a classification scheme of MLPs that has been proposed in Ref.~\citenum{behler2021machine}, but we note that any classification scheme is necessarily limited in view of the large variety of methods, and different schemes might equally be employed.

The PES is a high-dimensional function depending on all degrees of freedom of all atoms in the system. Therefore, its accurate approximation is a challenging task. First generation MLPs~\cite{behler2021four} were restricted to low-dimensional systems containing only a small number of atoms or freezing most degrees of freedom to restrict the complexity of the  PES~\cite{blank1995neural}. MLPs were generalized to high-dimensional, i.e., condensed systems by the introduction of high-dimensional neural network potentials (HDNNP)\cite{P1174}. In this first example of a second-generation (2G) MLP it was proposed~\cite{P1174} to construct the total potential energy $E$ as a sum of local environment-dependent atomic energy contributions $E_i$,
\begin{equation}
E = \sum_{i=1}^{N_\mathrm{atoms}} E_i,
\label{eq:e_decom}
\end{equation}
where $N_\mathrm{atoms}$ is the number of atoms in the system. 

The key to enable the ansatz of  Eq.~\ref{eq:e_decom} has been the introduction of many-body descriptors confined to the local environment around each atom defined by a cutoff radius. A cutoff between 6\,\AA~and 10\,\AA~is often a reasonable choice for many condensed-phase systems. Suitable descriptors must fulfill the physically mandatory translational, rotational and permutational invariances of the potential energy. Since the advent of second-generation MLPs many different types of descriptors meeting these requirements have been introduced~\cite{P2882,P3885,P5803,P6960}. Since no bonding patterns need to be specified, and atoms are allowed to enter and leave the local environments in molecular dynamics simulations, such descriptor-based MLPs can describe the making and breaking of chemical bonds in the same way as the underlying reference method. 

Nowadays, many MLPs are available that employ the locality ansatz of 2G MLPs. Similarly to HDNNPs, Gaussian Approximation Potentials (GAP),\cite{P2630} use predefined descriptors to approximate the PES. However, instead of atomic neural networks, kernel regression is employed for representing the atomic energies. Deep Potential Molecular Dynamics (DeePMD)\cite{P5596} shares a similar structure with HDNNPs but seeks to alleviate some of the difficulties of finding an optimal set of descriptors for representing the local atomic environment~\cite{P5076}. The DeePot-SE descriptor used in DeePMD contains an additional embedding network, which allows for data-driven optimization of the representation of the local atomic environment~\cite{zeng_deepmd-kit_2023}. Moment Tensor Potentials (MTPs)\cite{shapeev2016moment} and the Atomic Cluster Expansion (ACE)\cite{drautz2019atomic} represent the local atomic environment based on a body-order expansion, using bases of symmetric nonlinear functions, and regularized linear regression for training\cite{behler2021machine, cheng_cartesian_2024}. In particular, ACE introduced a framework for constructing the many-body basis at a constant cost per basis function irrespective of the order of the expansion, which mitigates the otherwise unfavorable scaling of many-body expansions\cite{drautz2019atomic}. 

Message Passing Neural Networks (MPNN)\cite{P5368} represent the atomic configurations as graphs where atoms form the nodes, which are connected to neighboring nodes within a cutoff distance by 
edges. Here, the representation of the atomic environments is learned by message passing operations along edges, where atoms exchange information about their state, which includes information on their position, chemical element and learnable features\cite{cheng_cartesian_2024, batatia_design_2025}. Messages can be iteratively propagated along edges in message passing steps, which leads to an effective enlarging of the receptive field of an atom beyond the cutoff distance\cite{cheng_cartesian_2024, batatia_design_2025}. Increasing the number of message passing steps is not equivalent to increasing the cutoff distance because only information from neighboring atoms, which can be reached in the specified number of steps along the graph, contributes to an atom\cite{batatia_design_2025}. This represents a sparsification of the interactions contributing to an atomic environment. Therefore, MPNNs can be considered semi-local\cite{batatia_design_2025}. After the message passing phase, the resulting state of an atom is mapped onto the atomic energy contribution (cf. Eq.~\ref{eq:e_decom}) using a learnable function, e.g., a neural network\cite{batatia_design_2025}. During training, both the representation as well as the mapping between the representation and the atomic energy contribution are optimized simultaneously\cite{batatia_design_2025}. MLPs that belong to this category include DTNN~\cite{schutt2017quantum}, SchNet~\cite{P5366}, AimNet~\cite{P5817}, PaiNN\cite{schutt2021equivariant}, NequIP~\cite{batzner20223}, GemNet\cite{gasteiger_gemnet_2024}, Equiformer\cite{liao_equiformer_2023}, Allegro~\cite{musaelian2023learning}, MACE~\cite{P6572} and GRACE~\cite{P6787}. Since in practical applications the number of message passing steps is rather small, MPNNs can also be considered as second-generation MLPs with finite range.

Embedding chemical element information in the representation of the atomic environment in MPNNs 
extends the applicability of
MLPs to 
large chemical spaces by keeping the number of descriptors
compact\cite{chen_graph_2019} and eliminating the need for element-specific predefined descriptors, 
whose number scales combinatorially with the number of elements in the system~\cite{behler2021machine}. It should be noted that element embedding is not exclusive to MPNN architectures, but can also be achieved using different methods such as the embedding network in DeePMD\cite{zeng_deepmd-kit_2023}. Recently, MPNN architectures have been used to train MLPs on databases containing large structural and chemical diversity, such as the Materials Project\cite{jain_commentary_2013} and the Open Materials 2024 (OMat24)\cite{barroso2024open} dataset,
with the aim to develop \textit{foundation models} that are transferable to a broad range of systems for materials discovery\cite{takamoto_towards_2022, batatia2023foundation}. Pre-trained MLPs which belong to this category include M3GNet\cite{chen_graph_2019}, PreFerred Potential\cite{takamoto_towards_2022}, MACE-MP0\cite{batatia2023foundation}, SevenNet\cite{park_scalable_2024}, MatterSim\cite{yang2024mattersim} and Orb\cite{neumann2024orbfastscalableneural, rhodes2025orbv3atomisticsimulationscale}. These developments are discussed further in Section~\ref{sec:Conclusions and Outlook}.

The introduction of a distance cutoff -- or a finite number of message passing steps -- necessarily limits the accuracy of 2G MLPs for systems where long-range interactions, primarily electrostatics but also dispersion interactions, are important. For such systems, an explicit treatment of long-range interactions can become necessary to achieve an accurate MLP~\cite{artrith2011high,yue_when_2021, anstine_machine_2023}. Long-range electrostatics have been incorporated into MLPs by learning local environment-dependent atomic charges\cite{P2391,artrith2011high,P3132,P5577,P5313,P5885,P5372,gastegger2017machine} or maximally localized Wannier centers\cite{zhang_deep_2022} from which the interactions can be determined using Coulomb's law or Ewald summation\cite{ewald_berechnung_1921}. In a similar spirit, long-range dispersion interactions have been incorporated by learning local environment-dependent Hirshfeld volumes\cite{muhli_machine_2021,tu_neural_2023} and exchange-hole moments~\cite{tu_neural_2023} for the Tkatchenko-Scheffler\cite{tkatchenko_accurate_2009} and exchange-hole dipole moment\cite{becke_exchange-hole_2005, becke_exchange-hole_2007} models, respectively. MLPs that include an explicit long-range interaction term determined from local environment-dependent atomic properties like, e.g., charges, can be classified as third-generation (3G) MLPs~\cite{behler2021four}.

\begin{figure}[!ht]
\centering
\includegraphics[width=0.4\textwidth]{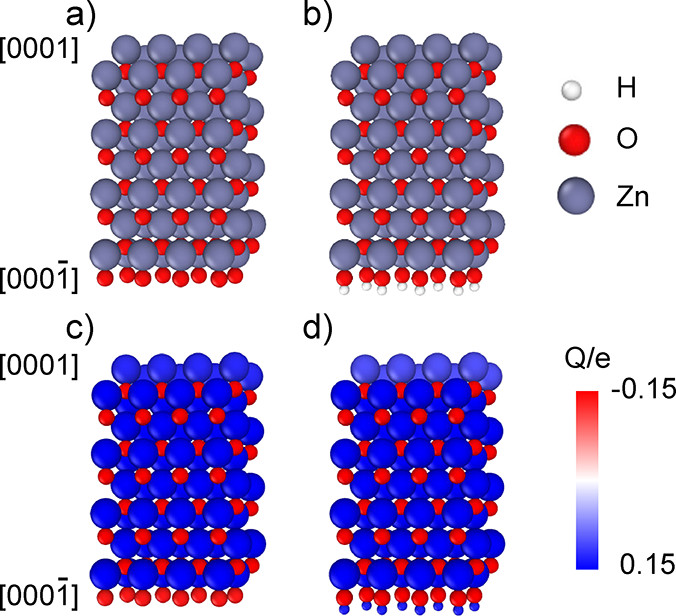}
  \caption{
  Side views of a polar Zn oxide slab structure (a) and (b) and the corresponding DFT Hirshfeld charge distributions (c) and (d) obtained using the PBE\cite{perdew_generalized_1996} functional. The [0001] surface of both slabs is Zn-terminated and the [$000\bar{1}$] surface is oxygen terminated. In (b) and (d), an additional hydrogen layer is attached to the [$000\bar{1}$] surface. The additional hydrogen layer alters the global charge redistribution, which leads to lower surface charges in (d) compared to (c) and thus in a reduced dipole moment of the slab. 
  Reproduced with permission from Behler, Chem. Rev. \textbf{121}, 16, 10037-10072 (2021). Copyright 2021 American Chemical Society\cite{behler2021four}.}
  \label{fig:ZnO_4G}
\end{figure}

A remaining limitation of the MLPs discussed so far is the inability to describe nonlocal phenomena like long-range charge transfer, which requires global information beyond the local atomic environment.
This is demonstrated for a polar zinc oxide (ZnO) slab in Figure~\ref{fig:ZnO_4G}, which shows the structure and Hirshfeld charge distribution of all atoms in the system~\cite{behler2021four}. The [0001] surface of the slab is Zn-terminated, while the [$000\bar{1}$] surface is oxygen-terminated, which gives rise to a net dipole moment in the system. On the right-hand side, in Figures~\ref{fig:ZnO_4G}(b) and (d), an additional hydrogen layer is attached to the oxygen-terminated surface, which globally alters the charge distribution in the system as can be seen by comparing the Hirshfeld charges at both surfaces for the slab with and without the additional hydrogen layer. The Zn-terminated surface does not obtain any information about the presence or absence of the hydrogen layer in second- and third-generation MLPs, hence the same charges and energies are assigned to these atoms in both cases such that the different systems cannot be distinguished by local MLPs. 
MLPs that can describe global phenonmena like long-range charge transfer define the fourth-generation (4G) of MLPs. Methods such as CENT\cite{ghasemi2015interatomic,khajehpasha2022cent2}, 4G-HDNNPs\cite{ko2021fourth}, PANNA\cite{shaidu2024incorporating} and kQEq\cite{staacke_kernel_2022} introduce a global charge equilibration scheme into the model for this purpose. Similarly, SpookyNet\cite{unke2021spookynet} is a fourth-generation MPNN architecture that includes an additional message passing operation between atoms independent of their distance, which introduces non-locality into the representation of the atomic environment.

Though long-range effects and nonlocality become important in cases such as the polar ZnO slab discussed above, to date the vast majority of MLPs used for oxide-water interfaces belong to the second generation. Moreover, the majority of existing studies on oxide-water interfaces, discussed in Section~\ref{sec:Applications}, employ DeePMD or HDNNPs. Given the growing number of emerging MLPs showing excellent performance, we anticipate that the diversity of models used for studying oxide-water interfaces will substantially increase in the coming years, including more frequent use of third- and fourth-generation approaches.

Among the key challenges in training MLPs for oxide-water interfaces is the simultaneous presence of bulk solid, bulk liquid, and interfacial regions -- each with distinct structural and electronic characteristics that must be accurately captured by the potential. The configuration space is therefore large, and becomes larger still if defects, surface reconstructions, multiple surface terminations or even doping are considered. Furthermore, many interfacial processes of interest, such as PT reactions, are rare events governed by subtle fluctuations in the hydrogen-bonding network. Ensuring that the relevant local environments associated with these processes are adequately represented in the training set, while maintaining a manageable training set size, requires careful data selection strategies. Active learning -- where candidate data points are iteratively selected based on some measure of model uncertainty or information content -- has become increasingly important for this purpose. One common approach is query-by-committee\cite{seung1992query, smith2018less, schran2020committee, schran2021machine}, where disagreement over an ensemble of models is used as a proxy for model uncertainty. In addition, enhanced sampling techniques such as metadynamics or umbrella sampling may be coupled with active learning to ensure rare but important events are well represented in the training data\cite{andrade2020free, yang2022using, zeng2023mechanistic, kobayashi2024long, tokita2025free}.

Once trained, the quality of an MLP is typically assessed by the root mean square error (RMSE) of energies and forces over an unseen test set. 
Though useful, such metrics are strongly influenced by the test set composition, which may not fully capture the configurational diversity encountered during large-scale molecular dynamics simulations. 
As a result, low RMSEs do not necessarily guarantee simulation stability or reliable reproduction of physical observables~\cite{fu2022forces}, and should be complemented by validation based on physical observables with respect to the reference method (e.g., lattice parameters, radial distribution functions, interfacial water density profiles, etc.) and long-term simulation stability. 
Finally, an important consideration is the choice of DFT method used to generate training data, as oxide-water interfaces -- particularly those involving transition metal oxides -- pose significant challenges for DFT. This point is discussed further in Section~\ref{sec:Conclusions and Outlook}.

\section{Applications to oxide-water interfaces}
\label{sec:Applications}
\subsection{Water dissociation and proton transfer}\label{sec:Water dissociation and proton transfer}

Interfacial water at metal oxide surfaces exhibits very different properties compared to the bulk liquid. For instance, the water molecules show enhanced reactivity, leading to a dynamic environment with frequent PT reactions. Adsorbed water molecules may dissociate by donating a proton to a surface oxygen site, leaving behind a -- often also adsorbed -- hydroxide ion that may undergo further PT reactions with other water molecules, or recombine with protons to form water. 
Such processes significantly modify the local water structure, charge distribution, and hydrogen-bond network at the interface, thereby influencing a broad range of properties possibly relevant in catalysis, where metal oxides are, e.g., promising candidates as anodes for the oxygen evolution reaction\cite{burke2015oxygen}. 
Right from the beginning, understanding the extent and mechanisms of water dissociation on specific surface facets has therefore been a major focus of many studies on oxide-water interfaces employing MLPs.

\subsubsection{Zinc oxide-water interfaces}\label{sec:Zinc oxide-water interfaces}

\begin{figure}[!ht]
\centering
\includegraphics[width=0.5\textwidth]{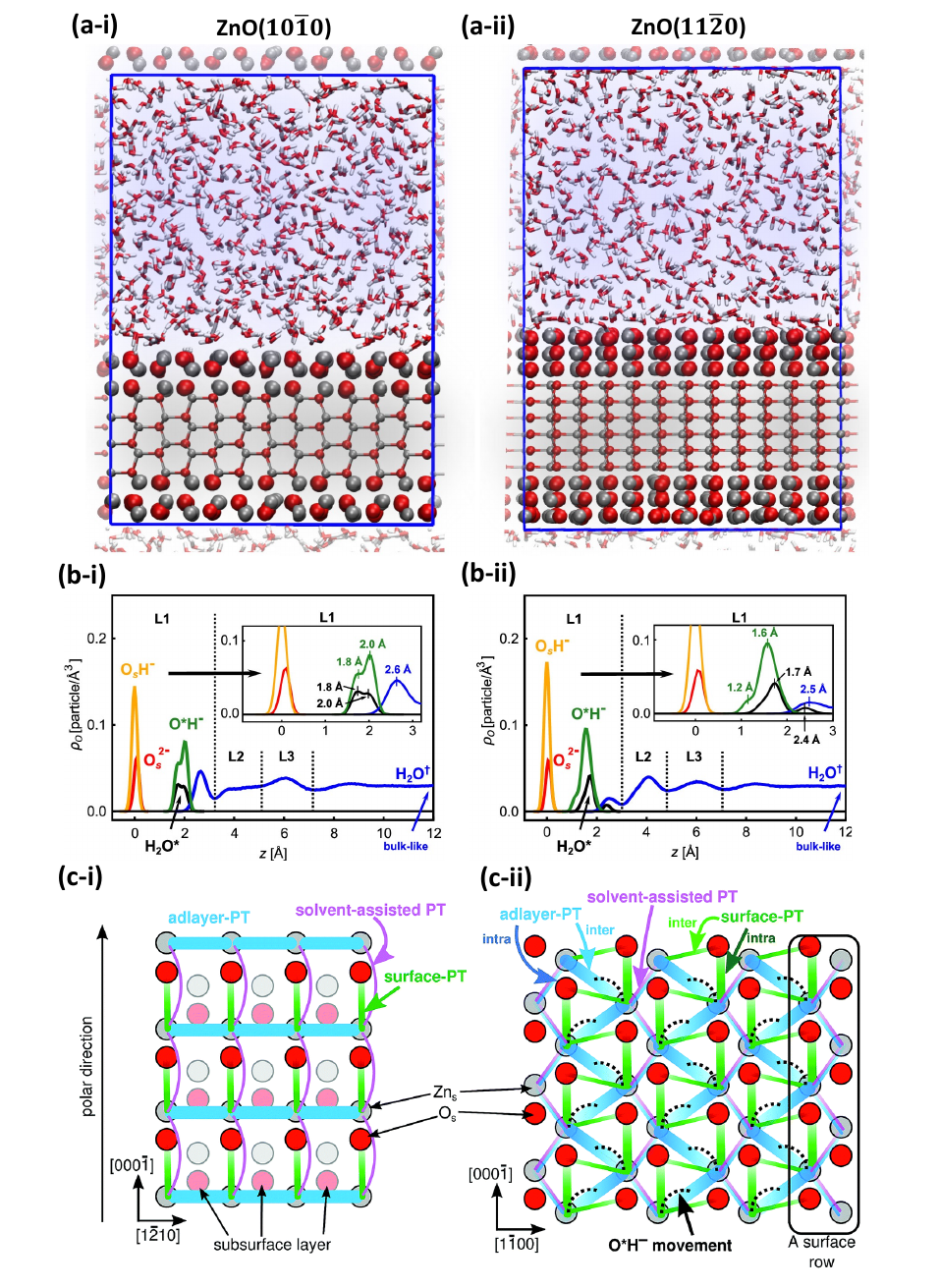}
  \caption{(a) Side views of atomistic models of the (a-i) ZnO($10\bar{1}0$)- and (a-ii)  ZnO($11\bar{2}0$)-water interfaces. (b) Density profiles of oxygen species along the surface normal direction for the (b-i) ZnO($10\bar{1}0$)- and (b-ii)  ZnO($11\bar{2}0$)-water interface models. 
  (c) Top views of the long-range PT networks on the (c-i) ZnO($10\bar{1}0$) and (c-ii)  ZnO($11\bar{2}0$) facets. Thicker lines represent PT reactions with higher rates, while color gradients from dark to light indicate larger barriers for the forward reaction.
  (a) and (b) 
  Adapted with permission from Quaranta \emph{et al.}, J. Phys. Chem. C \textbf{123}, 1293-1304 (2019).
  Copyright 2018 American Chemical Society\cite{quaranta2018structure}.
  (c) 
  Adapted with permission from Hellstr\"om \emph{et al.} Chem. Sci., \textbf{10}, 1232-1243 (2019). Copyright 2019 Authors, licensed under a Creative Commons Attribution 3.0 Unported License\cite{hellstrom2019one}.}
  \label{fig:ZnO}
\end{figure}

The first MLP-based simulations of oxide-water interfaces were conducted by Quaranta \emph{et al.}\cite{quaranta2017proton}, who employed an HDNNP
to investigate water dissociation and PT reactions at the nonpolar ZnO(10$\bar{1}$0)-water interface. 
Two dominant PT mechanisms were observed: (i) PT between an adsorbed water molecule (H$_2$O$^*$) and a neighboring surface oxygen site (O$_s$), denoted surface-PT, and (ii) PT between an adsorbed water molecule and an adjacent hydroxide ion (O$^*$H$^-$), denoted adlayer-PT. 
Water dissociation via the surface-PT mechanism was found to be rather extensive, with on average 71~\% of interfacial water molecules in the dissociated state.

The relative rates of adlayer- and surface-PT reactions were investigated by computing the associated free energy surfaces (FESs) based on a one-dimensional PT coordinate\cite{tuckerman2002nature}, revealing a substantially lower barrier for adlayer-PT compared to surface-PT and thus a faster reaction rate. A key finding was that PT rates for both types of reaction are governed by a predominant presolvation mechanism, wherein thermal fluctuations in the local hydrogen-bonding environment transiently lower the free energy barrier for PT events. Overall, the obtained results are in very good agreement with previous AIMD studies~\cite{tocci2014solvent}. 

The HDNNP was later extended to study the nonpolar ZnO($11\bar{2}0$)-water interface \cite{quaranta2018structure}, where the same PT mechanisms were observed, with comparable PT time scales, 
and a similar extent of dissociated water (76~\%). Atomistic models of the two interface systems are shown in Figure~\ref{fig:ZnO}(a), each containing several thousand atoms. Figure~\ref{fig:ZnO}(b) shows density profiles of different oxygen species as a function of distance from the surface, illustrating the presence of O$^*$H$^-$ (green) and O$_s$H$^-$ (orange) species resulting from water dissociation. Bulk-like behaviour of water is recovered at about 10~$\text{\AA}$ from the surface.

Protons at interfaces may diffuse over large distance via a Grotthuss-like mechanism\cite{marx2006proton}. 
This was investigated at the two ZnO-water interfaces by tracking the mean squared displacement of proton hole centers (PHCs)\cite{hellstrom2019one}, representing the positions of ``missing protons'', i.e., unprotonated surface oxygen sites ($\mathrm{O_s^{2-}}$) or adsorbed hydroxide ions ($\mathrm{O^*H^-}$) along with the corresponding surface Zn ion. 
The resulting PHC diffusion networks, shown in Figures~\ref{fig:ZnO}(c-i) and (c-ii), reveal a strong facet dependence. At ZnO(10$\bar{1}$0), PHC diffusion is primarily one-dimensional, dominated by adlayer-PT along the nonpolar $[1\bar{2}10]$ direction (blue, Figure~\ref{fig:ZnO}(c-i)). Surface-PT along the polar $[0001]$ direction acts as a dead end (green, Figure~\ref{fig:ZnO}(c-i)), and long-range PT along this direction may only proceed via a rare solvent-assisted mechanism (purple, Figure~\ref{fig:ZnO}(c-i)). 

In contrast, ZnO($11\bar{2}0$) supports two-dimensional PHC diffusion, with contributions from both surface- and adlayer-PT. These processes can be further classified as intra-row and inter-row PT, distinguishing transfers within the same surface row from those between adjacent rows, where a surface row is indicated by the black box in Figure~\ref{fig:ZnO}(c-ii). Surface-PT predominantly involves intra-row transport, while adlayer-PT facilitates inter-row transport via diagonal pathways. In summary, it has been found that the oxide surface geometry has a strong impact on PT mechanisms along the surface, which thus could be potentially controlled by the design of suitable surface structures.

\subsubsection{Titania-water interfaces}\label{sec:Titania-water interfaces}

TiO$_2$-water interfaces have received considerable attention due to their importance in photocatalysis\cite{fujishima1972electrochemical, linsebigler1995photocatalysis, schneider2014understanding, guo2019fundamentals}. The extent of dissociated water molecules at such interfaces is of particular interest because surface hydroxyl groups are thought to act as trapping sites for photoinduced charge carriers, thereby influencing photocatalytic activity\cite{anpo1985esr, szczepankiewicz2000infrared}.

\begin{figure}[!ht]
\centering
\includegraphics[width=0.45\textwidth]{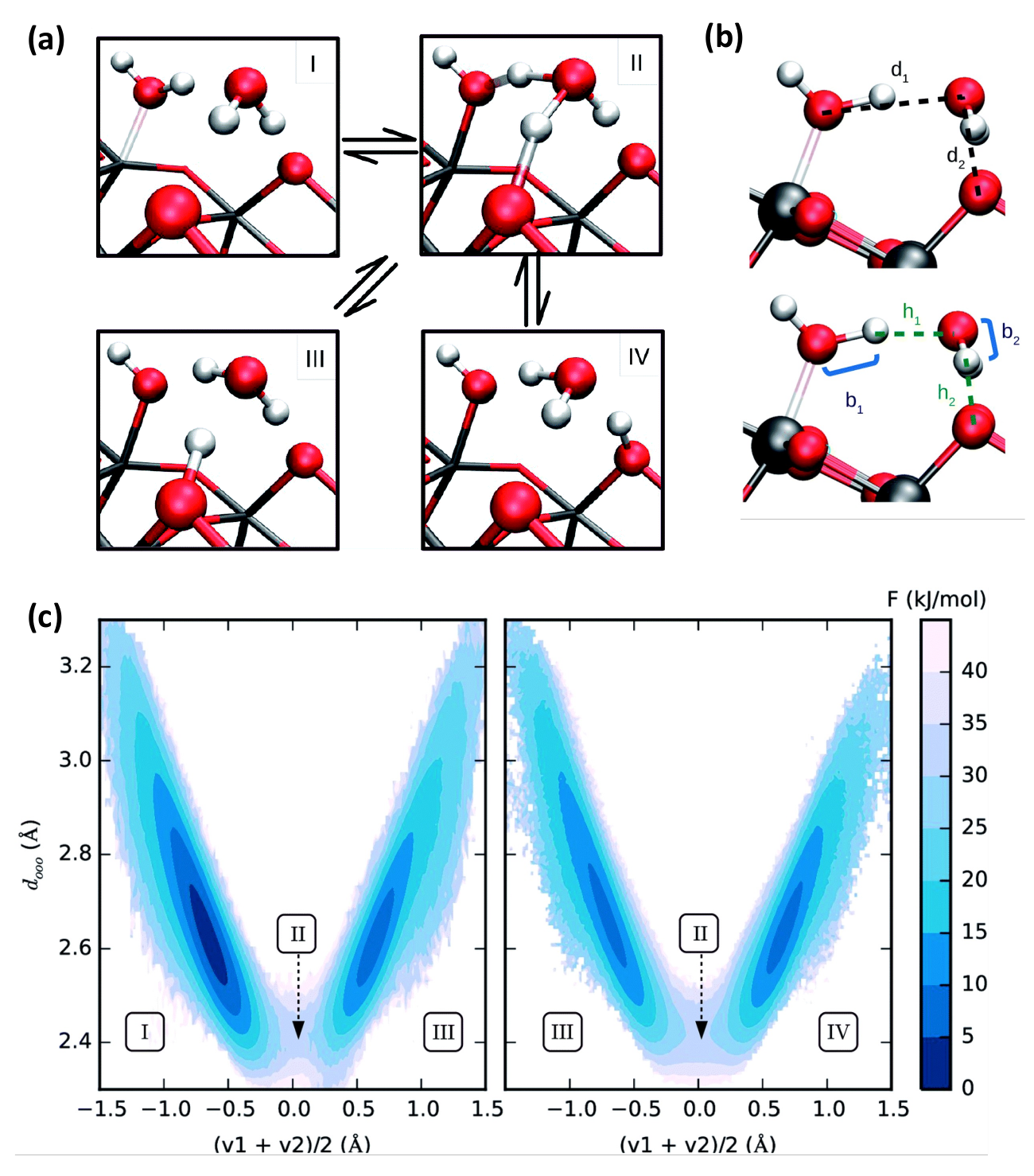}
  \caption{(a) Mechanistic pathways for water dissociation and PT at the anatase TiO$_2$(101)-water interface. (b) Interatomic distances used to define the collective variables $d_{ooo} = (d_1 + d_2) / 2$ and $(v_1 + v_2) / 2$ for PT reactions, where $v_i = b_i - h_i$. (c) FES for different pathways illustrated in panel (a) obtained from umbrella sampling simulations. 
  Adapted with permission from Andrade \emph{et al.}, Chem. Sci., \textbf{11}, 2335-2341 (2020). Copyright 2020 Authors, licensed under a Creative Commons Attribution-NonCommercial 3.0 Unported License\cite{andrade2020free}.}
  \label{fig:Andrade2020_FES}
\end{figure}

The extent and mechanism of water dissociation at the anatase TiO$_2$(101)-water interface was investigated by Andrade \emph{et al.}\cite{andrade2020free},  
where
DeePMD simulations revealed that on average only 5.6~\% of interfacial water molecules are dissociated. Water dissociation was found to proceed via a solvent-assisted concerted PT mechanism (in contrast to the direct surface-PT mechanism discussed above for ZnO). The mechanism is illustrated in Figure~\ref{fig:Andrade2020_FES}(a) (pathways I $\rightarrow$ II $\rightarrow$ III and I $\rightarrow$ II $\rightarrow$ IV). 
Similarly, solvent-assisted PT was observed between adjacent surface oxygen sites (pathway III $\rightarrow$ II $\rightarrow$ IV). 
FESs for the different pathways in Figure~\ref{fig:Andrade2020_FES}(a) were computed from umbrella sampling simulations (Figure~\ref{fig:Andrade2020_FES}(c)), 
showing water dissociation to be endergonic
by 8.0~kJ/mol
with a barrier of 32 kJ/mol. This is in line with the low extent of dissociated water observed in the unbiased simulations. 

DeePMD simulations of water confined within anatase(101) nanopores\cite{kwon2024confinement} revealed that the mechanism of water dissociation remained consistent across pore sizes from 5 to 20~$\text{\AA}$ (following the I $\rightarrow$ II $\rightarrow$ III pathway in Figure\ref{fig:Andrade2020_FES}(a)). However, strong confinement within a 5~$\text{\AA}$ pore was found to lower free energy barriers, leading to faster PT, due to a contraction of oxygen-oxygen distances within the hydrogen-bond network. Confinement was also found to restrict the water oxygen mobility while enabling relatively fast and directional proton diffusion. For the unconfined system, hydrogen mobility was primarily driven by the collective motion of water molecules.

The presence of formic and acetic acid adsorption at the anatase TiO$_2$(101)-water interface was also investigated 
using DeePMD\cite{raman2024long}, revealing that interfacial water structuring strongly depends on acid coverage. At high acid coverage, a transition from the bidentate to a monodentate configuration of the adsorbed acid was observed, accompanied by the adsorption of interfacial water molecules onto the vacated Ti sites.

In addition to anatase(101), several MLPs have been reported for the rutile TiO$_2$(110)-water interface\cite{schran2021machine, zhuang2022resolving, wen2023water}.
DFT studies have shown that water adsorption energy on this surface exhibits oscillations with respect to slab thickness\cite{harris2004molecular, liu2010structure}, arising from variations in O$_{2p}$-Ti$_{3d}$ hybridization induced by the presence of a central layer in odd-numbered slabs\cite{bredow2004electronic}. Such studies were, however, restricted to monolayer water coverage or less. 
Zhuang \emph{et al.}\cite{zhuang2022resolving} computed the equilibrium fraction of dissociated water at the rutile TiO$_2$(110)-water interface as a function of slab thickness using PBE+D3-based DeePMD simulations, showing the presence of odd-even oscillations with respect to number of slab layers. These oscillations decayed with increasing slabs layers, with the water dissociation fraction converging to a value of 2~\%. Wen \emph{et al.}\cite{zhuang2022resolving} observed the same oscillations using a SCAN-based DeePMD potential, though a slightly larger extent of dissociation, 22~\%, was reported for thick slabs. This discrepancy is likely due to differences in the DFT method used for training, highlighting the sensitivity of simulation outcomes to the choice of reference method and the importance of careful validation of the underlying electronic structure approach.
We note that conducting simulations using sufficiently thick slabs to obtain converged interfacial water properties is only possible due to the efficiency of MLPs.

Zeng \emph{et al.}\cite{zeng2023mechanistic} investigated water dissociation across seven low-index TiO$_2$ surfaces through HDNNP-driven metadynamics simulations. 
Water dissociation was found to be thermodynamically favourable on anatase(100), anatase(110), rutile(001), and rutile(011), whereas anatase(101), rutile(100) and rutile(110) favoured molecular adsorption. 
Furthermore, PT mechanisms across the different surfaces were classified using a sparsified kernel principal component analysis (kPCA) based on geometric features such as interatomic distances and displacement coordinates relevant to PT events, which enabled the automated identification of distinct PT pathways.

\subsubsection{Ceria-water interfaces}\label{sec:Ceria-water interfaces}

Kobayashi \emph{et al.}\cite{kobayashi2024long} investigated long-range PT and hydroxide ion transport at the CeO$_2$(110)- and (111)-water interfaces using DeePMD. 
A wide variety of PT mechanisms was observed, including the adlayer-PT (APT) and surface-PT (referred to as here as surface proton formation/recombination, or SPF/R) mechanisms discussed above for ZnO-water interfaces\cite{quaranta2017proton}, 
as well as a third mechanism involving PT between surface hydroxyl groups and unprotonated surface oxygen ions (denoted SPT). These mechanisms were further categorized into direct (I) and solvent-assisted (II) types, with the latter proceeding via intermediate $\mathrm{OH^-}$ or $\mathrm{H_3O^+}$ species in the solvent. Direct migration of $\mathrm{OH^-}$ ions within the adlayer was also observed, referred to as adlayer hydroxide transfer (AHT). 
Long-range PT dynamics was analysed in terms of Semi-Markov state models, where PT is treated as a stochastic process between discrete states, enabling quantification of reaction rates.

\begin{figure}[!ht]
\centering
\includegraphics[width=0.5\textwidth]{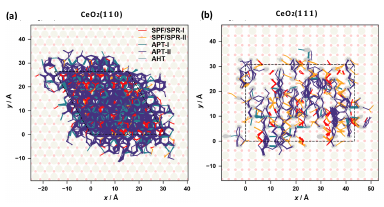}
  \caption{Long-range PT networks at (a) the CeO$_2$(110)-water interface and (b) the CeO$_2$(111)-water interface. 
  Different types of PT mechanisms are indicated with different colours; surface proton formation/recombination (SPF/R), adlayer PT (APT) -- where the types I and II indicate direct and solvent assisted mechanisms -- and adlayer hydroxide transfer (AHT).
  Adapted with permision from Kobayashi \emph{et al.}, Chem. Sci., \textbf{15}, 6816-6832 (2024). Copyright 2024 Authors, licensed under a Creative Commons Attribution 3.0 Unported License\cite{kobayashi2024long}.}
  \label{fig:CeO2}
\end{figure}

PHC trajectories at the CeO$_2$(110) and CeO$_2$(111)-water interfaces are shown in Figures~\ref{fig:CeO2}(a) and (b), respectively.
The CeO$_2$(111) interface shows much higher PT activity than the CeO$_2$(110) interface, which can be compared with the strong facet dependence of long-range PT observed at ZnO-water interfaces\cite{hellstrom2019one} (cf. Figure~\ref{fig:ZnO}(c)).

\subsubsection{Acid-base properties}\label{sec:Acid-base properties}

The propensity for water to dissociate near an oxide surface is intimately linked to the acidity constants ($\mathrm{pK_a}$) of the exposed surface sites, which quantify the equilibrium between protonated and deprotonated species as a function of pH. An important related quantity is the point of zero proton charge, $\mathrm{pH_{PZC}}$, which denotes the pH at which the net surface charge is zero and which can be computed from the $\mathrm{pK_a}$ of surface sites. 
For a given surface deprotonation reaction, the $\mathrm{pK_a}$ is related to the free energy change, $\Delta A$, of the reaction by the expression

\begin{equation}
    \mathrm{pK_a} = \frac{1}{\ln(10)}\frac{\Delta A}{k_BT},
    \label{eq:pKa}
\end{equation}
where $k_B$ is the Boltzmann constant and $T$ the temperature.

Knowledge of surface $\mathrm{pK_a}$ values can be extremely valuable for rationalising the properties of solid-liquid interfaces. 
For instance, in 
DeePMD simulations of the rutile TiO$2$(110)-water interface, Zhuang and Cheng\cite{zhuang2023deciphering} reported the unexpected deprotonation of surface Ti$_{5\text{c}}$OH$^-$ species to form Ti$_{5\text{c}}$O$^{2-}$ groups which persisted over $\sim 100$~ps. 
This behavior was rationalized by computing the $\mathrm{pK_a}$ values for Ti$_{5\text{c}}$OH$^-$ and Ti$_{5\text{c}}$O$^{2-}$ using AIMD-based thermodynamic integration combined with free energy perturbation theory\cite{sulpizi2008acidity, cheng2009redox}.
These studies revealed that the $\mathrm{pK_a}$ of Ti$_{5\text{c}}$OH$^-$ (5.61) is in fact lower than that of Ti$_{5\text{c}}$OH$_2$ (8.32).
This reversal of the expected acidity order was attributed to structural relaxation effects; specifically, a shortening of the Ti-O bond and a transition of the Ti coordination from octahedral to pyramidal in Ti$_{5\text{c}}$O$^{2-}$. 

Recently, Schienbein \emph{et al.} reported the calculation of $\mathrm{pK_a}$ values for surface oxygen sites at the BiVO$_4$-water interface using HDNNP-driven thermodynamic integration and free energy perturbation theory. 
Notably, a single MLP was able to represent the PESs of the initial (``protonated'') state, the final (``deprotonated'') state, and mixed states corresponding to a linear combination of the initial and final states\cite{schienbein2024data}. 

Another route for calculating $\mathrm{pK_a}$ values is enhanced sampling simulations. 
Raman and Selloni\cite{raman2023acid} employed DeePMD-driven well-tempered metadynamics simulations 
to compute the FESs for interfacial deprotonation reactions at the IrO$2$(110)-water interface, enabling the evaluation of surface $\mathrm{pK_a}$ values according to Eq.~\ref{eq:pKa}. Specifically, they determined $\mathrm{pK_{a,br}} = 3.6$ for the deprotonation of a surface bridging oxygen (O$\mathrm{_{br}}$) and $\mathrm{pK_{a,cus}} = 3.2$ for the deprotonation of a terminal water molecule adsorbed to a coordinatively unsaturated Ir site ($\mathrm{Ir_{cus}}$). These values allowed for estimation of the point of zero proton charge as $\mathrm{pH_{PZC}}=\frac{1}{2}(\mathrm{pK_{a,br}} + \mathrm{pK_{a,cus}}) = 3.4$, in good agreement with the experimental estimate of $\simeq 3$\cite{kosmulski2016isoelectric}.
Moreover, since the overall water dissociation reaction can be expressed in terms of these two individual deprotonation steps, the corresponding free energy change could be computed as $\Delta A_\text{diss} = \ln(10) k_B T [\mathrm{pK_{a,cus} - pK_{a,br}}] = \ln(10) k_B T [\Delta \mathrm{pK_a}] = -1.8$~$\mathrm{kJ/mol}$.
This negative value indicates that water dissociation at the IrO$_2$ surface is thermodynamically favorable, consistent with the high fraction of dissociated water molecules observed in unbiased DeePMD simulations ($\simeq$80~\%).

As noted in Ref.~\citenum{jia2024water}, where PT reactions were studied at the rutile SnO$_2$-water interface with DeePMD, it is insightful to compare trends across the series of isostructural rutile oxides, including TiO$_2$, SnO$_2$, and IrO$_2$. 
The acidity difference between adsorbed water and protonated bridge oxygen atoms, $\Delta \mathrm{pK_a}$, is notably larger for TiO$_2$(110)\cite{cheng2015reductive} than for SnO$_2$(110)\cite{jia2020computing} and IrO$_2$(110)\cite{raman2023acid}. This correlates with a significantly lower water dissociation fraction ($\alpha$) for TiO$_2$ ($\sim 0.02$)\cite{zhuang2022resolving}, compared to SnO$_2$ ($\sim 0.63$)\cite{jia2024water} and IrO$_2$ ($\sim 0.80$)\cite{raman2023acid} observed in MLP-based MD simulations.
Moreover, $\mathrm{pK_a}$ values have also been computed for aqueous silicate oligomers\cite{roy2024learning} using umbrella sampling simulations driven by PaiNN\cite{schutt2021equivariant}, an equivariant message-passing architecture.

\subsection{Defects and dynamic surfaces}\label{sec:Defects and dynamic surfaces}

In section~\ref{sec:Water dissociation and proton transfer}, we focused on how water behaves near ideal, defect-free oxide surfaces. 
In practice, however, oxide–water interfaces are seldom found in such a pristine state. Defects such as steps, kinks, and vacancies are practically unavoidable in experimental settings and in nature, and can drastically influence processes such as catalysis and corrosion\cite{campbell2005oxygen, diebold2010oxide, yu2020asymmetric}. 
Moreover, initially pristine surfaces under aqueous conditions may undergo substantial reconstruction, altering their morphology and reactivity. Metal ions may also adsorb to or desorb from metal-oxide surfaces in contact with a liquid phase -- an important phenomenon in geochemistry\cite{davis1986geochemical}. In this section, we review MLP-based studies that address these deviations from perfect bulk-truncated single-crystal surfaces, ranging from the role of defects to dynamic changes in surface structure via reconstruction, diffusion or adsorption processes.

\subsubsection{Defects at oxide-water interfaces}\label{sec:Defects at oxide-water interfaces}

\begin{figure*}[!ht]
\centering
\includegraphics[width=0.8\textwidth]{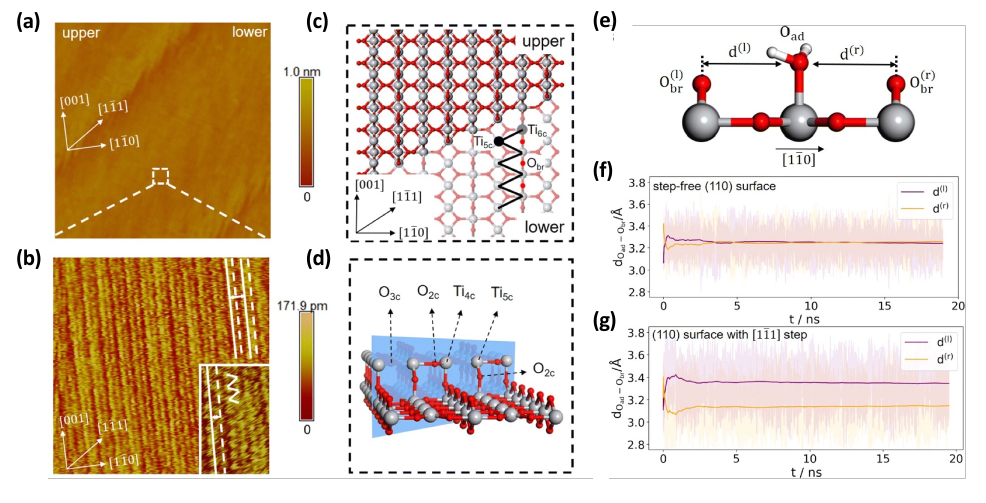}
  \caption{(a) EC-STM image ($150\times150~\mathrm{nm}^2$) of rutile TiO$_2$(110) in 0.1 M HClO$_4$. (b) Magnified image ($10\times10~\mathrm{nm}^2$) corresponding to the white dashed square in panel (a), with the inset showing a magnified area of $3\times5~\mathrm{nm}^2$. (c) Top view of the atomistic model of the stepped rutile TiO$_2$(110) surface. (d) Side view of the atomistic model showing the step edge. (e) Horizontal distances, $d^{(l)}$ and $d^{(r)}$, between an adsorbed water molecule and neighbouring briding oxygen atoms (O$_\text{br}$). (f), (g) Computed distances $d^{(l)}$ and $d^{(r)}$ over a molecular dynamics trajectory for the step-free and stepped surface models, respectively. The accumulated average of $d^{(l)}$ and $d^{(r)}$ are shown in bold lines. 
  Adapted with permission from Sun \emph{et al.}, Chem. Sci., \textbf{15}, 12264-12269 (2024). Copyright 2024 Authors, licensed under a Creative Commons Attribution-NonCommercial 3.0 Unported License\cite{sun2024step}.}
  \label{fig:TiO2_step_defect}
\end{figure*}

The impact of step defects at rutile TiO$_2$(110) on interfacial water structuring was recently highlighted by the 
observation of a double-row pattern in interfacial water using electrochemical scanning tunneling microscopy (EC-STM), shown in Figures~\ref{fig:TiO2_step_defect}(a) and~(b)\cite{sun2024step}.
DeePMD simulations 
reported in the same study
demonstrated that the double-row pattern could only be rationalized when a surface [$1\bar{1}1$] step was explicitly included in the model. Top and side views of the model surface step structure are shown in Figures~\ref{fig:TiO2_step_defect}(c) and (d), respectively. The time evolution of distances between an adsorbed water oxygen and its neighboring bridging surface oxygen atoms (denoted $d^{(l)}$ and $d^{(r)}$ in Figure~\ref{fig:TiO2_step_defect}(e)) are shown in Figures~\ref{fig:TiO2_step_defect}(f) and (g) for the pristine and stepped surface models, respectively. 
At the pristine surface, adsorbed water remains symmetrically distributed, as evidenced by the similar averaged values of $d^{(l)}$ and $d^{(r)}$. In contrast, water molecules at the stepped surface shift asymmetrically, forming a double-row pattern of oxygen density with spacing in excellent agreement with the experiments reported in the same study. 
This shift originates from the influence of the step edge, where water molecules near the step experience asymmetric interactions leading to a lateral displacement, which propagates across the terrace through the hydrogen bond network.

An extreme example of structural disorder is found in amorphous systems, which effectively represent a high density of defects. Ding and Selloni investigated the amorphous TiO$_2$-water interface using 
DeePMD simulations\cite{ding2023modeling}, showing that 
water near the amorphous surface exhibits markedly greater structural disorder than at crystalline facets, lacking the characteristic peaks in the water oxygen density profile. Notably, the diffusion coefficient of interfacial water molecules was found to be an order of magnitude higher than on crystalline surfaces. 

Nakanishi \emph{et al.}\cite{nakanishi2025defect} recently reported 
HDNNP simulations of the prisine ZrO$_2$-water interface and the Zr$_7$O$_8$N$_4$-water interface, where the latter represents a ZrO$_2$ system containing a large number of oxygen vacancies (V$_\mathrm{O}$) and oxygen-to-nitrogen substitutions (N$_\mathrm{O}$).
The equilibrium distribution of the defect sites within the oxide slab was obtained using MLP-driven replica exchange Monte Carlo simulations\cite{artrith2017efficient, kasamatsu2022facilitating, hoshino2023probing}, revealing that V$_\mathrm{O}$ defects preferentially arrange into uniform columns perpendicular to the surface, while N$_\mathrm{O}$ defects primarily segregate into the third anion layer from the surface, with very few appearing at the surface itself. 
The extent of dissociated water was found to be greater at the pristine ZrO$_2$ surface compared to Zr$_7$O$_8$N$_4$, attributed to the larger amount of surface O sites available to accept protons.
Moreover, water at the Zr$_7$O$_8$N$_4$ surface was found to preferentially adsorb on Zr atoms adjacent to V$_\mathrm{O}$ sites rather than directly at the vacancy sites, leaving the vacancies largely unoccupied.  
This suggests that V$_\mathrm{O}$ sites remain free from water ``poisoning", which has implications for catalytic activity\cite{doi2007zirconium, yin2013enhancement}. 
PT mechanisms were subsequently investigated using a protocol similar to that applied to ZnO-water interfaces\cite{quaranta2017proton, hellstrom2019one} (Section~\ref{sec:Zinc oxide-water interfaces}),
showing distinct presolvation mechanisms for surface-PT on pristine ZrO$_2$ and Zr$_7$O$_8$N$_4$, but no presolvation mechanism for adlayer-PT in either system.
Furthermore, long-range PT was found to proceed much more efficiently on the pristine ZrO$_2$ surface than on Zr$_7$O$_8$N$_4$, due to disruption of PT networks at V$_\mathrm{O}$ sites.

Lee and Lee\cite{lee2025machine} explored reconstructed terminations of the BiVO$_4$(010) surface using simulated annealing simulations\cite{kresse2003complex} driven by a GAP to identify low-energy structures across a broad range of stoichiometries, including Bi- and V-rich surfaces exhibiting diverse bonding environments with varying oxygen concentrations. 
The electrochemical stability of selected terminations was assessed by constructing Pourbaix diagrams using hybrid DFT, followed by AIMD simulations with explicit water to characterize interfacial water adsorption and dissociation.
While the stoichiometric surface showed no water dissociation, several reconstructed surfaces exhibited spontaneous dissociation, particularly at exposed Bi sites.

The water interface of the magnetite(001) subsurface vacancy model (SCV)\cite{bliem2014subsurface}, characterized by an ordered arrangement of subsurface vacancies and interstitials, has recently been investigated with HDNNP-based MD simulations\cite{romano2024structure}. The simulations identified new stable water adsorption configurations at low coverage and revealed anisotropic water diffusion along the surface iron rows.

\subsubsection{Dynamic surfaces and ion desorption/adsorption}\label{sec:Dynamic surfaces and ion desorption/adsorption}

Various processes involving dynamic motion or rearrangement of atoms belonging to the solid oxide at its water interface have been reported using MLPs. At the water interface of tobermorite (a calcium-silicate-hydrate found in cement\cite{brunauer1962tobermorite}), 
HDNNP simulations revealed lateral diffusion of Ca$^{2+}$ ions within the surface plane at the interface with water\cite{kobayashi2021machine}. 
Similarly, 
DeePMD simulations of the muscovite mica-water interface\cite{raman2024ab, raman2024insights} showed that K$^+$ ions diffuse via hopping between surface cavities, facilitated by interactions with interfacial water that transiently lift ions away from the surface.
To assess the likelihood of full ion desorption, well-tempered metadynamics simulations were employed to compute the associated FES for interface models with varying arrangements of K$^+$ ions.
In all cases, full K$^+$ desorption was found to be highly unfavorable, with a free energy cost of about 34~kJ/mol\cite{raman2024insights}.

Ion dissolution was also investigated at the dicalcium silicate ($\beta$-Ca$_2$SiO$_4$(100))-water interface\cite{li2023unravelling} using DeePMD-driven metadynamics simulations 
to map out the FES for Ca$^{2+}$ dissolution across temperatures from 300 and 500~K. Dissolution was found to be thermodynamically favourable at all temperatures considered, though the reaction mechanism was found to vary, shifting from an associative ligand-exchange pathway at lower temperatures to a dissociative mechanism at higher temperatures. Dissolution time scales were estimated using frequency-adaptive metadynamics for kinetic rate calculations\cite{wang2018frequency} and were found to range from hundreds of seconds at 300~K to nanoseconds at 500~K.

\begin{figure}[!ht]
\centering
\includegraphics[width=0.45\textwidth]{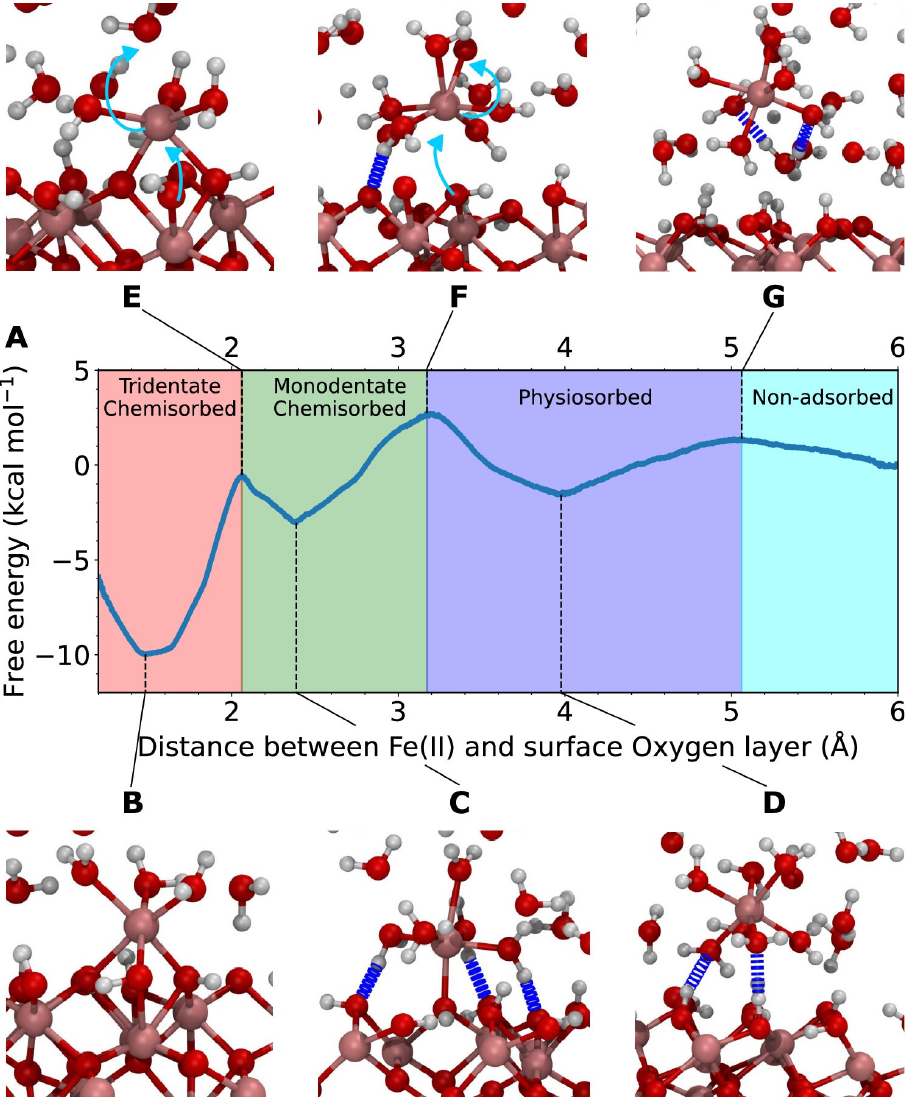}
  \caption{(A) FES for Fe$^{2+}$ adsorption on hematite(001) as a function of ion-surface distance along the surface normal direction from umbrella sampling simulations.
  Representative snapshots corresponding to local minima (B--D) and maxima (E--G) of the FES are shown in the insets. Reproduced with permission from Joll \emph{et al.}, J. Phys. Chem. Lett., \textbf{16}, 848-856 (2025). Copyright 2025 Authors, licensed under a Creative Commons Attribution (CC BY) License\cite{joll2025mechanism}.}
  \label{fig:Fe_chemisorption}
\end{figure}

Recently, Joll \emph{et al.} investigated the chemisorption mechanism of aqueous Fe$^{2+}$ ions onto a hematite surface\cite{joll2025mechanism}, building on a committee-HDNNP model developed for the 
hydroxyl-terminated hematite(001)-water interface\cite{schienbein2022nanosecond}.
Umbrella sampling simulations were first employed to investigate the mechanism and rate of water ligand exchange in the Fe$^{2+}$-hexaaqua complex, which showed a dissociative mechanism where the departure of a coordinating water molecule is followed by the uptake of another solvent water molecule. A similar mechanism has been found also for the exchange of ligand water in NaOH solutions, which is assisted by PT to neighboring hydroxide ions~\cite{P5126}.
Subsequently, the FES for Fe$^{2+}$ adsorption onto the hematite(001) surface was computed using umbrella sampling, reproduced in Figure~\ref{fig:Fe_chemisorption}, revealing an overall exergonic process with four distinct states; nonadsorbed, physisorbed, monodentate chemisorbed, and tridentate chemisorbed.
Representative snapshots corresponding to the stationary points of the FES are shown in Figure~\ref{fig:Fe_chemisorption}(b)-(g). 
The transition from physisorbed to chemisorbed states involves the partial rupture of Fe–O bonds with first-shell water ligands and the sequential formation of Fe–O bonds with surface hydroxyl groups, culminating in a highly stable tridentate chemisorbed configuration approximately 7.1~kcal/mol lower in free energy compared to the monodentate chemisorbed structure.
Notably, earlier force field studies\cite{kerisit2015computational} found the Fe$^{2+}$ chemisorption process to be endergonic, due in part to the rigidity of surface hydroxyl groups oriented along the surface normal, which hindered Fe$^{2+}$ ion coordination with surface oxygen atoms. This highlights how force fields can yield starkly different thermodynamic predictions compared to \textit{ab initio}-based MLPs.

Dynamic surface reconstruction due to strong interaction with water has been reported in DeepMD simulations of an Mxene-water interface\cite{hou2023unraveling}.
MXenes are a class of two-dimensional transition metal carbide/nitrides; here, an oxygen-terminated V$_2$CO$_2$ slab was considered. 
Water molecules, with their oxygen atom oriented toward the surface, were found to interact strongly with subsurface V atoms, occasionally pulling them outward to form V–O bonds. Subsequent proton release from the adsorbed water molecule led to the formation of stable surface vanadium oxide groups, with the released protons forming hydronium ions. 
Over time, the accumulation of surface vanadium oxide groups and solvated protons was found to inhibit further oxidation, resulting in an exponential decay in the oxidation rate, also observed in experiments for similar MXene systems\cite{zhang2017oxidation, huang2019hydrolysis, doo2021mechanism}.
The role of hydrated protons in oxidation inhibition was further explored by varying the proton concentration in nanoconfined water between MXene sheets\cite{hou2024proton}. Higher proton concentrations were found to reduce the number of oxidation-active water molecules by altering their orientation near the surface. Moreover, at low proton concentrations, a room-temperature hexagonal ice-like phase 
was observed within the nanoconfined water layer, stabilized by a combination of in-plane hydrogen bonding and strong interactions with the 
surface.

Similarly, dynamic surface reconstruction has been observed in simulations of Co-Fe-based perovskite oxide systems in contact with liquid water\cite{li2024molecular} using VASP's on-the-fly MLPs based on GAPs\cite{jinnouchi2019fly}.
The simulations revealed that water interacts strongly with the oxide surface, leading to termination-dependent restructuring, including the exchange of lattice oxygen with water, the formation of surface peroxo species, and in some cases, the release of molecular O$_2$ into solution. These processes primarily involved oxygen atoms originally bonded to Co, reflecting the weaker Co-O bond compared to Fe-O.

\subsection{Beyond energies and forces}\label{sec:Beyond energies and forces}

MLPs provide an efficient mapping between atomic structure and energy, with forces and stresses available as analytic derivatives of this energy function. While this framework has greatly expanded our ability to simulate large-scale systems, a drawback of conventional MLPs is the loss of explicit electronic-structure information, which is critical for properties that directly depend on the underlying electron density. 
For instance, vibrational spectra such as IR, Raman and SFG are determined by selection rules linked to properties of the total charge distribution, i.e., dipole moments and polarizabilities, which determine the spectral intensity of different vibrational modes in the spectrum. Likewise, modeling electrochemical phenomena often requires knowledge of electrostatic potentials depending on the charge distribution. Consequently, employing machine learning to recover relevant electronic structure information is an active area of research\cite{handley2009dynamically, artrith2011high, grisafi2018transferable, yeo2019pattern, christensen2019operators, eckhoff2020predicting, litman2020temperature, zhang2020deep, moreno2021machine, gastegger2021machine, shao2022finite, ceriotti2022beyond, hafizi2023ultrafast, zhang2023universal, joll2024machine, sowa2024bond, jana2024learning, sommers2020raman, schienbein2023spectroscopy}. Here, we discuss recent studies that, in addition to using MLPs, employ complementary ML models to predict key electronic-structure properties relevant for modeling oxide–water interfaces. 

\subsubsection{Oxidation states in lithium manganese oxide}\label{sec:Oxidation states in lithium manganese oxide}

Eckhoff and Behler investigated different pristine LiMn$_2$O$_4$-water-interface orientations and terminations using an HDNNP.\cite{eckhoff2021insights} LiMn$_2$O$_4$ shows electrocatalytic activity for the oxygen evolution reaction (OER), which is the rate limiting step of the electrolysis of water.\cite{cady_tuning_2015, Tahir2017} The manganese ions coexist in the oxidation states IV and high-spin (hs)-III, necessitating the use of a hybrid density functional to obtain the correct electronic structure.\cite{eckhoff_hybrid_2020} This valency of the manganese ions has been shown to relate with LiMn$_2$O$_4$'s OER activity\cite{cady_tuning_2015}. Even though HDNNPs do not include explicit information on the electronic structure, Eckhoff \emph{et al.}\cite{eckhoff2020predicting} showed in a related study that this information can be learned implicitly by an HDNNP and the oxidation state distribution can be recovered from the local atomic environment using an additional high-dimensional neural network trained on atomic spins (HDNNS). Making use of the presence or absence of Jahn-Teller distorted Mn coordination by oxygen atoms, HDNNS can predict the spin value of the manganese ions from the representation of their local environment, similar to the prediction of environment-dependent charges in 3G HDNNPs\cite{artrith2011high}. This can be used to classify the oxidation states ($S_\mathrm{Mn(IV)}=1.5\,\hbar$ and $S_\mathrm{hs-Mn(III)}=2.0\,\hbar$) of the manganese ions in the system,\cite{eckhoff2020predicting} 
and Eckhoff and Behler employed an HDNNS in tandem with an HDNNP to investigate the oxidation state distribution at the LiMn$_2$O$_4$-water interface, which they found to differ significantly from the bulk solid depending on the surface orientation and termination.\cite{eckhoff2021insights}

\subsubsection{Vibrational spectroscopy}\label{sec:Vibrational spectroscopy}

Vibrational spectra such as IR, Raman and SFG can be computed from the Fourier transform of correlation functions in terms of dipole moments and/or polarizability tensors. 
Machine learning these quantities to enable efficient computational spectroscopy is becoming increasingly common, as demonstrated by recent studies on aqueous systems\cite{sommers2020raman, schienbein2023spectroscopy, shepherd2021efficient, litman2023fully, kapil2024first, o2023elucidating, du2024revealing, berrens2025molecular}, gas-phase molecules\cite{gastegger2017machine, pracht2024efficient, sowa2024bond} and bulk crystals\cite{berger2024polarizability}. 

\begin{figure}[!ht]
\centering
\includegraphics[width=0.5\textwidth]{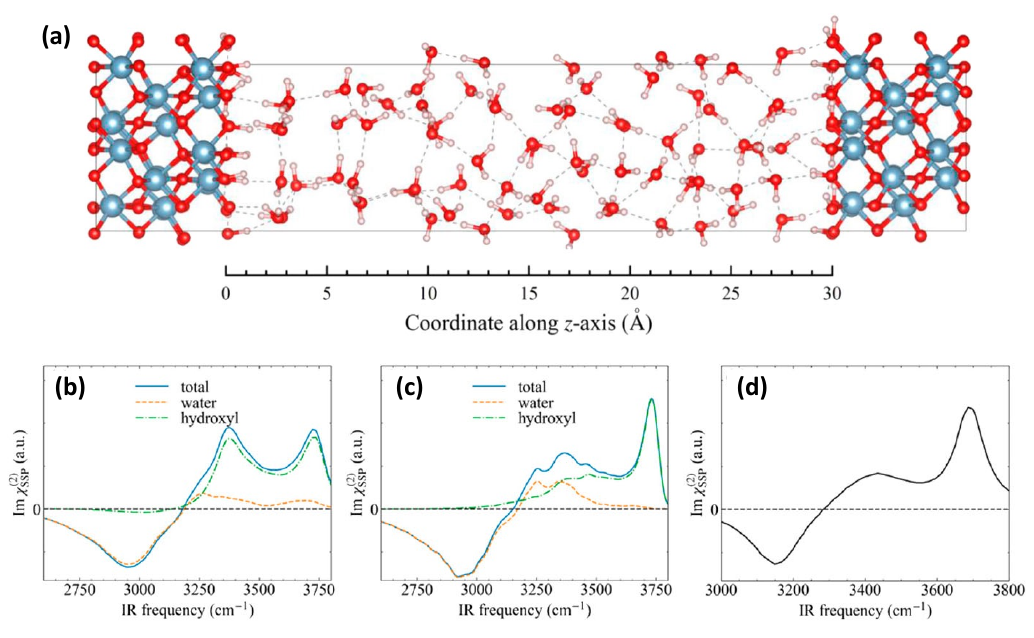}
  \caption{(a) Atomistic model of the $\alpha$-Al$_2$O$_3$(0001)-water interface. (b), (c) and (d) show SFG spectra computed using the $\alpha$-$\mu$ approach, the ssVVCF approach and experimental data\cite{zhang2008structures}, respectively. The theoretical spectra are decomposed into contributions from water (orange, dashed) and surface hydroxyl groups (green, dashed). 
  Adapted from Du \emph{et al.} J. Chem. Phys. \textbf{161}, 124702 (2024), with the permission of AIP publishing. Copyright 2024 AIP Publishing\cite{du2024revealing}.
  The data in panel (d) are adapted with permission from Zhang \emph{et al.} J. Am. Chem. Soc. \textbf{130}(24), 7686-7694 (2008). Copyright 2008 American Chemical Society\cite{zhang2008structures}. The sign of the experimental data was reversed for consistency with the system geometry in the simulations.} 
  \label{fig:Al2O3_spectroscopy}
\end{figure}

In particular, SFG spectroscopy is a powerful tool for characterizing interfacial systems due to its inherent sensitivity to regions of broken centrosymmetry, enabling selective probing of the interface\cite{shen1989surface}.
Du \emph{et al.}\cite{du2024revealing} computed SFG spectra for the hydrogen-terminated $\alpha$-Al$_2$O$_3$(0001)-water interface using DeePMD in combination with Deep Wannier models\cite{zhang2020deep,sommers2020raman} to predict molecular dipole moments and polarizability tensors.
An atomistic model of the interface is shown in Figure~\ref{fig:Al2O3_spectroscopy}(a). 

In general, the SFG spectrum is related to the Fourier transform of the dipole moment-polarizability correlation function\cite{mukamel1995principles, nagata2013water} (referred to here as the $\mu$-$\alpha$ approach). 
The molecular dipole moments of water molecules and surface hydroxyl groups can be computed as\cite{sharma2005intermolecular}

\begin{equation}\label{eq:dipole_water}
    \mu_{\text{H}_2\text{O}, i} = e\left(6\mathbf{r}^i_\text{O} + \mathbf{r}^i_{\text{H}, 1} + \mathbf{r}^i_{\text{H}, 2} - 2\sum_{j=1}^{4}\mathbf{r}_{\text{W},j}\right)
\end{equation}
and
\begin{equation}\label{eq:dipole_oh}
    \mu_{\text{OH}, i} = e\left(6\mathbf{r}^i_\text{O} + \mathbf{r}^i_\text{H} + \frac{1}{2}\mathbf{r}^i_{\text{Al}, 1} + \frac{1}{2}\mathbf{r}^i_{\text{Al}, 2} - 2\sum_{j=1}^{4}\mathbf{r}_{\text{W},j}\right),
\end{equation}
where $e$ is the elementary charge, $\mathbf{r}_\text{H}$ and $\mathbf{r}_\text{O}$ are the coordinates of the H and O atomic nuclei, respectively, and $\mathbf{r}_{\text{Al}, 1}$ and $\mathbf{r}_{\text{Al}, 2}$ are the coordinates of the two Al nuclei bonded to the surface hydroxyl group. The index $i$ runs over molecules and $\mathbf{r}_{\text{W}, j}$ represents the $j^\text{th}$ Wannier center. 
Two separate Deep Wannier models were trained; one to directly predict Wannier centroids, i.e., the average over Wannier centers assigned to a given atom, and another to predict their response to applied electric fields, from which polarizability tensors can be calculated.
Additionally, the authors computed SFG spectra via the approximate surface-specific velocity–velocity correlation function (ssVVCF) approach\cite{ohto2015toward}, which is based on the velocity autocorrelation function and parameterized expressions for the dipole moment and polarisability\cite{corcelli2005infrared, auer2008ir}.  IR and Raman spectra were also computed. 

The SFG spectra computed using the $\mu$-$\alpha$ approach and the ssVVCF approach are shown in Figures~\ref{fig:Al2O3_spectroscopy}(b) and (c), respectively, while experimental data from Ref.~\citenum{zhang2008structures} is shown in Figure~\ref{fig:Al2O3_spectroscopy}(d). 
Both computational approaches reproduce the main experimental features and allow for a
decomposition into contributions from interfacial water (orange) and surface hydroxyl groups (green). 
However, discrepancies between the $\mu$-$\alpha$ and ssVVCF approaches are evident, particularly in the intensity of the first positive peak around 3350~cm$^{-1}$. 
This was attributed to the use of a parameterized polarizability in the ssVVCF method, which cannot resolve different components of the polarizability tensor. 
Discrepancies between the $\mu$-$\alpha$ spectrum and experiment may stem from the known limitations of the PBE+D3 functional, the neglect of nuclear quantum effects (NQEs), and differences in the surface protonation state.

In another study\cite{o2023elucidating}, water structuring on anatase TiO$_2$(101) was investigated at different water coverage levels up to two mono-layers
based on computed IR spectra from DeePMD simulations, using a separate DeePMD model to predict the system's total dipole moment. Frequency shifts and intensity variations in the spectrum as a function of water coverage could largely reproduce trends from experiments also reported in that study and could be linked to underlying changes in hydrogen bonding and molecular orientations with increasing coverage.

We note that even without access to dipole moments or polarizability tensors, analysis of the vibrational signatures at oxide-water interfaces may be highly informative. 
For example, HDNNPs were used to compute the anharmonic OH stretching vibrational spectrum at the ZnO($10\bar{1}0$)–water interface\cite{quaranta2018maximally}, showing that the surface perturbs OH stretching frequencies up to 4~$\text{\AA}$.
At the hematite(100)-water interface\cite{schienbein2022nanosecond}, analysis of the vibrational density of states (VDOS) of H atoms in first-layer water from HDNNP simulations revealed a red shift in the OH stretching frequencies for bonds oriented perpendicular to the surface,
indicating a weakening of the intramolecular OH bond due to strong hydrogen bonding interactions with the surface.
Finally, in DeePMD simulations of thermal transport at the anatase TiO$_2$–water interface\cite{li2023thermal}, analysis of the interfacial VDOS showed that water dissociation has a strong impact on interfacial thermal conductance. 

\subsubsection{The electrical double layer at the titania-electrolyte interface}\label{sec:The electrical double layer at the titania-electrolyte interface}

Electric double layers (EDLs) dictate the electrochemical environment at solid-electrolyte interfaces\cite{parsons1990electrical, gonella2021water, li2023electric}. 
They are commonly described within the framework of mean-field models, such as the Gouy-Chapman-Stern (GCS) model\cite{kinraide1994use, oldham2008gouy, bard2022electrochemical}, which assumes a continuum dielectric representation of water and a uniformly distributed surface charge, neglecting specific atomic-scale structure and dynamics.

To assess the validity of this model for the anatase TiO$_2$(101)-electrolyte interface, Zhang \emph{et al.}\cite{zhang2024molecular} conducted MD simulations over a range of electrolyte solutions, employing the Deep Potential Long Range (DPLR) model\cite{zhang_deep_2022}, which incorporates long-range electrostatics. 
This is an example of a 3G MLP (see section~\ref{sec:Machine learning potentials}).
The inclusion of long-range interactions was found to be necessary for describing oxide-electrolyte interfaces to avoid unphysical ion distributions with unbalanced charge in the bulk electrolyte.
Furthermore, Deep Wannier models\cite{zhang2020deep,sommers2020raman} were used to predict Wannier centroids associated with each atom, enabling the calculation of the electrostatic potential profile by solving Poisson's equation.

\begin{figure}[!ht]
\centering
\includegraphics[width=0.5\textwidth]{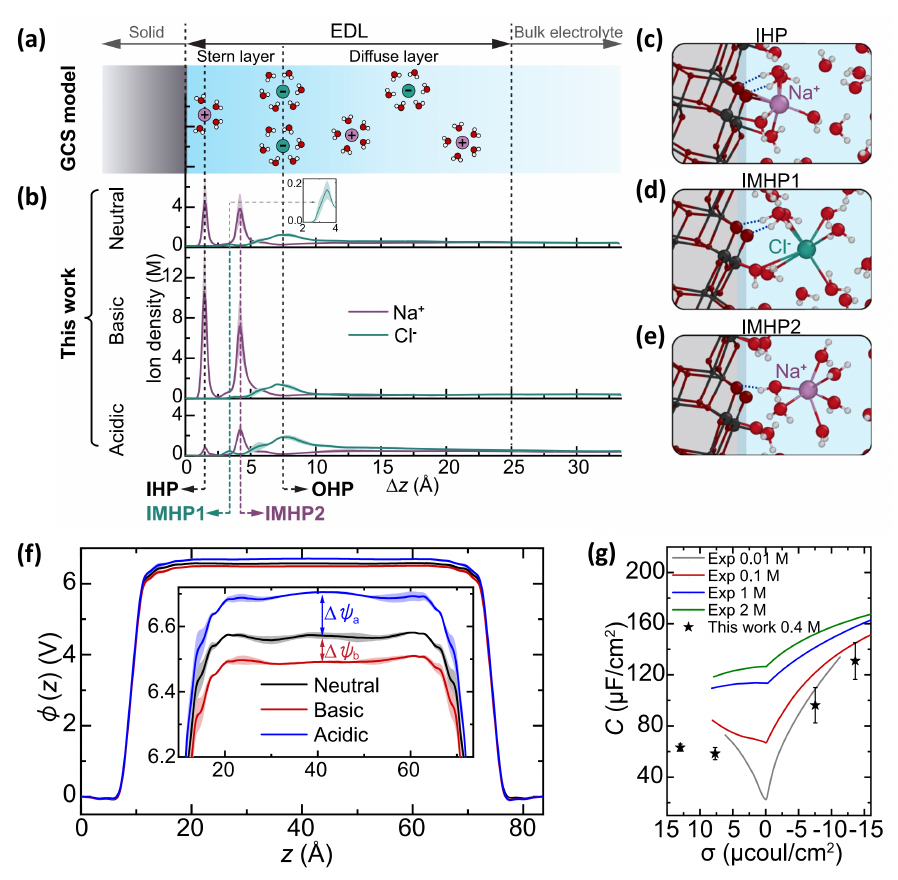}
  \caption{(a) Schematic of the Gouy-Chapman-Stern (GCS) model of the electrical double layer (EDL). (b) Ion distributions obtained from MD simulations for the neutral, basic and acidic solutions. The inner and outer Helmholtz planes (IHP and OHP, respectively) are indicated by dashed black lines. Additional ion density peaks observed in the simulations are denoted as intermediate Helmholtz planes (IMHP1 and IMHP2). (c-e) Representative snapshots of structures corresponding to the IHP, IMHP1 and IMHP2. (f) Macro-averaged electrostatic potential $\phi$ along the surface normal z-direction for neutral (black), basic (red) and acidic (blue) solutions. 
  (g) Computed interfacial capacitance (star symbols) and experimental values\cite{berube1968adsorption1,be1968adsorption2} (solid lines) as a function of the surface charge density. Adapted with permission from Zhang \emph{et al.}, Nat. Commun. 2024, \textbf{15}, 10270 (2024). Copyright 2024 Springer Nature\cite{zhang2024molecular}.}
  \label{fig:TiO2_electrolyte}
\end{figure}

Electrolyte solutions covering a range of pH conditions were considered. 
Under neutral conditions (0.4~M NaCl), the TiO$_2$ surface remained charge neutral, with balanced proton and hydroxide adsorption arising from water dissociation. However, under basic (0.4~M NaCl + 0.2~M NaOH) and acidic (0.4~M NaCl + 0.2~M HCl) conditions, the surface became negatively and positively charged, respectively, due to preferential adsorption of hydroxide ions or protons.

A schematic of the GCS model is shown in Figure~\ref{fig:TiO2_electrolyte}(a) and the ion distributions obtained in the simulations under different pH conditions are shown in Figure~\ref{fig:TiO2_electrolyte}(b).
Compared to the GCS model, the simulations revealed two additional peaks within the Stern layer, termed intermediate Helmholtz planes (IMHPs), which correspond to semi-adsorbed Na$^+$ and Cl$^-$ ions (Figures~\ref{fig:TiO2_electrolyte}(d)-(e)). Notably, ion structuring was observed even under neutral conditions where the net surface charge is zero, due to the influence of electronegative surface oxygen atoms which attract Na$^+$ ions.

Electrostatic potential profiles under neutral, basic and acidic conditions are shown in Figure~\ref{fig:TiO2_electrolyte}(f) in black, red and blue, respectively. By analyzing the changes in surface charge density ($\sigma$) and in potential drop between the bulk solid and liquid regions ($\psi$) upon introducing basic or acidic conditions, the differential capacitance could be estimated as $C = \frac{d\sigma}{d\psi} \approx \frac{\Delta\sigma}{\Delta\psi}$,
shown in Figure~\ref{fig:TiO2_electrolyte}(g) alongside experimental measurements as a function of surface charge for different background salt concentrations\cite{berube1968adsorption1,be1968adsorption2}. The experimental trend of increasing capacitance with decreasing surface charge is well-reproduced by the simulations. This was attributed to the strong affinity of Na$^+$ ions to electronegative surface oxygen atoms, allowing closer approach compared to Cl$^-$ ions. In contrast, no such asymmetry is predicted by the GCS model. 
The simulations also revealed distinct surface charging mechanisms in acidic compared to basic conditions, reflecting differences in the ability of the EDL to screen surface charges.

\section{Conclusions and Outlook}\label{sec:Conclusions and Outlook}

The last few years have seen a rapid adoption of machine learning potentials as efficient surrogates for DFT-based atomistic simulations. Here, we have discussed their use in simulations of oxide-water interfaces, which are important for many technical applications, highlighting insights into the reactive behaviour of interfacial water, the influence of defects and dynamic surface processes. Additionally, we reviewed studies leveraging machine learning to recover electronic structural information, which is typically lost when training only on energies and forces, allowing for efficient access to spin state distributions, vibrational spectra, and electrostatic potentials. The low computational demand of MLPs compared to \textit{ab initio} calculations is key to enabling simulations of system sizes approaching realistic interfaces (i.e., with sufficient water to exhibit bulk-like behavior far from the surface, thick oxide slabs and the inclusion of complex defects), while also reaching long enough time scales to achieve statistical convergence of key properties, which is crucial for systems containing a liquid phase. Still, several important challenges remain to be solved to further improve the reliability of atomistic simulations of complex interfaces using machine learning. 

The accuracy of an MLP is inherently constrained by the choice of reference electronic structure method, namely the DFT exchange-correlation function. 
A key difficulty lies in achieving an accurate and consistent description of both the solid oxide and liquid water phases, as well as their interface.
Certain GGA functionals, when augmented with dispersion interactions, provide a good description of liquid water\cite{morawietz2016van}, but perform poorly on bulk transition metal oxides, significantly underestimating electronic band gaps\cite{robertson2006band, li2013density, pandey2017electronic}. 
More accurate hybrid functionals often improve the description of the bulk oxide by mitigating self-interaction errors, but these benefits do not necessarily carry over to the liquid phase\cite{gillan2016perspective}.
Alternatively, the DFT+U approach\cite{himmetoglu2014hubbard} can improve the description of localized electronic states in transition metal oxides, though its predictive power is sensitive to the choice of the U parameter, which lacks a universal prescription, eventually limiting the accuracy of the reference calculations to certain types of structures with consequences for the transferability of the potential.
Ultimately, it is important to benchmark the underlying density functional against experimental data as far as possible~\cite{harmon2020validating}.

Another source of uncertainty is the neglect of nuclear quantum effects (NQEs), such as zero point energy and tunnelling, in classical MD simulations where nuclei are treated as point particles evolving according to Newton's laws\cite{markland2018nuclear}. 
These effects can be captured by path integral molecular dynamics (PIMD)\cite{tuckerman2023statistical} and related approaches such as ring-polymer molecular dynamics (RPMD)\cite{habershon2013ring}, which map each quantum nucleus onto a set of ``beads'' that sample its quantum delocalization in an extended classical phase space.
The higher computational cost of path integral methods makes the efficiency of MLPs particularly attractive for large-scale simulations. 
Although the impact of NQEs at oxide-water interfaces is not yet well characterized, NQEs have been shown to have a strong impact across various systems involving hydrogen atoms\cite{ceriotti2016nuclear, hellstrom2018nuclear, yan2020nuclear, bocus2023nuclear, stolte2024nuclear} even at room temperature -- for instance, significantly enhancing water dissociation at metal surfaces\cite{litman2018decisive, cao2025quantum}. Incorporating NQEs into simulations of oxide-water interfaces is therefore an intriguing direction for future research. 

Another important research area is the modeling of electrified interfaces, which are central to applications such as electrocatalysis, energy storage, and sensing. The presence of electric fields at these interfaces can induce significant charge redistribution, drastically altering the local structure and dynamics\cite{che2018elucidating, futera2021water, gonella2021water, steinmann2021understanding}. 
As a result, properly accounting for charges, electrostatic potentials, and polarizability in MLPs becomes critical\cite{steinmann2021understanding} for future applications in this field. 
Although these electronic degrees of freedom can be captured using machine learning (see Section \ref{sec:Beyond energies and forces}), simulating electrochemical environments under constant electrode potential remains a key challenge, because it requires the exchange of charge with an external reservoir\cite{bonnet2012first, nielsen2015towards, melander2019grand, dufils2019simulating, deissenbeck2021dielectric}. Extending MLPs to support variable-charge or constant-potential frameworks, represents an important yet challenging area of method development.

Traditionally, MLPs are trained on data covering a limited chemical and configurational space -- for example, restricted to a single oxide-water interface, as in the studies reviewed here. As a result, a model trained on one oxide-water system cannot be readily applied to another.
While active learning can be employed to direct the extension of the configuration space covered by the reference data into new relevant regions, 
a significant change in chemical composition often requires developing a new potential from scratch.
In the last few years, advancements in MLP architectures (as discussed in Section~\ref{sec:Machine learning potentials}) and the growing availability of large and diverse databases has lead to the emergence of pre-trained MLPs which aim to cover the configuration space of a wide range of systems. 
Pre-trained models such as M3GNet\cite{chen_graph_2019} or MACE-MP-0\cite{batatia2023foundation}, trained on data from the Materials project\cite{jain_commentary_2013}, which includes extensive coverage of oxide systems,
are openly available and could facilitate simulations across a broad range of oxide-water systems. 
While promising, these models are a relatively recent development and data on their accuracy and applicability for interfacial systems is still limited. 
Batatia \emph{et al.}\cite{batatia2023foundation} evaluated the pre-trained MACE-MP0 potential on the SiO$_2$- and TiO$_2$-water interfaces  and found that it successfully reproduced key properties of interfacial water.
However, they note that the liquid phase is over-structured, due to the use of the PBE functional in the Materials Project. 
One way to address such limitations is by fine-tuning the pre-trained model's parameters using custom data at a higher level of theory, which is typically more data-efficient than training a new potential from scratch\cite{jacobs_practical_2025}.
The development of pre-trained potentials is a very active field and we expect rapid advances in the performance and applicability of available potentials. 

Finally, although most MLPs rely on the locality approximation of the atomic interactions (i.e., Eq.~\ref{eq:e_decom}), nonlocal effects can become significant under certain conditions. An example of relevance here is oxide doping, where the introduction of dopant atoms may induce global changes in the electronic structure\cite{ko2021fourth}.
In such scenarios, a single local model will be unable to account for changes in doping distribution or concentration, necessitating the use of more sophisticated models that incorporate nonlocal interactions in a more general way.

In summary, the application of MLPs to oxide-water interfaces has become an active and exciting yet challenging field of research. Apart from advanced applications pushing currently available methods to their limits, oxide-water interfaces have adopted a central role in stimulating further methodical developments by combining flexible and numerically accurate machine learning methods with physical concepts, which are essential to perform reliable computer simulations of complex interfaces.

\begin{acknowledgments}
JE and JB thank the Deutsche Forschungsgemeinschaft (DFG) for funding in CRC 1633 (C04, project number 510228793) 
and KNL and JB thank the DFG for funding in CRC 1073 (C03, project number 217133147). Moreover, we are grateful for support by the DFG under Germany's Excellence Strategy – EXC 2033 RESOLV (project number 390677874).
\end{acknowledgments}

\section*{Author Declarations}

The authors have no conflicts to disclose.

\section*{Data Availability Statement}

Data sharing is not applicable to this article as no new data were created or analyzed in this study.


\bibliography{literature}



\end{document}